\documentclass[pra,aps,twocolumn]{revtex4-1}
\usepackage{amsmath}
\usepackage{amssymb}
\usepackage{graphicx}
\usepackage{bbold}
\usepackage{csquotes}
\newcommand{\beqa}{\begin{eqnarray}}
\newcommand{\eeqa}{\end{eqnarray}}

\begin{document}
\title{$\mathcal{PT}$-breaking threshold in spatially asymmetric Aubry-Andre Harper models: hidden symmetry and topological states} 
\author{Andrew K. Harter, Tony E. Lee, Yogesh N. Joglekar}
\affiliation{Department of Physics, Indiana University Purdue University Indianapolis (IUPUI), 
Indianapolis, Indiana 46202, USA}

\date{\today}
\begin{abstract}
Aubry-Andre-Harper (AAH) lattice models, characterized by reflection-asymmetric, sinusoidally varying nearest-neighbor tunneling profile, are well-known for their topological properties. We consider the fate of such models in the presence of balanced gain and loss potentials $\pm i\gamma$ located at reflection-symmetric sites. We predict that these models have a finite $\mathcal{PT}$ breaking threshold only for {\it specific locations} of the gain-loss potential, and uncover a hidden symmetry that is instrumental to the finite threshold strength. We also show that the topological edge-states remain robust in the $\mathcal{PT}$-symmetry broken phase.   Our predictions substantially broaden the possible realizations of a $\mathcal{PT}$-symmetric system. 
\end{abstract}
\maketitle
%---------------------------------------------------------------------------------%

\section{Introduction}
\label{sec:intro}

When is the spectrum of a non-Hermitian Hamiltonian purely real or has only complex-conjugate pairs? Numerous authors have addressed this question, starting with Bender and coworkers who showed that (continuum) Hamiltonians invariant under combined parity- and time-reversal operations ($\mathcal{PT}$-symmetric) fit the bill~\cite{bender1,bender2,bender3}. Such Hamiltonians faithfully model open systems with balanced gain and loss, in which the parity operator ($\mathcal{P}$) exchanges the gain-region with the loss-region, whereas the time-reversal operator ($\mathcal{T}$) transforms a gain-region into a lossy region. Concurrent with their experimental realizations in coupled waveguides~\cite{expt1,expt2,expt3,uni1,uni2}, resonators~\cite{expt4}, microcavities~\cite{expt5}, and lasers~\cite{expt6,expt7,expt8,expt9}, discrete $\mathcal{PT}$ systems with a parity-symmetric tunneling term $H_0=\mathcal{P}H_0\mathcal{P}=H_0^\dagger$ and a balanced gain-loss potential $V=\mathcal{PT}V\mathcal{PT}\neq V^\dagger$ have been intensely studied in the past five years~\cite{bendix,song,znojil1,avadh,derek,longhi1}. In particular, site-dependent tunneling Hamiltonians, of interest for perfect-state-transfer and quantum computing~\cite{sougato}, have been theoretically~\cite{clint} and experimentally~\cite{faithH,albertoperuzzo,fabio} explored. All of these experimentally investigated systems have been subject to a the stringent constraint of a reflection-symmetric tunneling amplitude profile. 

Generically, the spectrum of a $\mathcal{PT}$-symmetric Hamiltonian $H=H_0+V$ is real when the strength $\gamma$ of the balanced gain-loss potential is smaller than a positive threshold $\gamma_{PT}$ set by $H_0$. The emergence of complex-conjugate eigenvalues at the exceptional point $\gamma=\gamma_{PT}$ is called $\mathcal{PT}$-symmetry breaking~\cite{kato,review}. When $\gamma>\gamma_{PT}$, the eigenfunctions with complex eigenvalues become increasingly asymmetrical~\cite{ltoc}. It has long been known that a purely real or complex-conjugate-pairs spectrum is equivalent to the existence of an antiunitary operator $A=U\mathcal{T}$ that commutes with the Hamiltonian $H$~\cite{mostafa,bender4,mannheim}. Thus, in principle, reflection-symmetry is not a necessary constraint, $U\neq\mathcal{P}$. Indeed there are several proposals, based on supersymmetric quantum mechanics, for continuum models where the complex potential $V(x)\neq V*(-x)$ is not reflection symmetric (or antisymmetric)~\cite{sqm1,sqm2}. Nonetheless, all experimental realizations of $\mathcal{PT}$-symmetric systems to-date have abided by the reflection-symmetry constraint. 

On a separate front, one-dimensional Aubry-Andre Harper (AAH) models have been extensively explored in recent years in the context of their experimental realizations in coupled waveguide arrays. Consider an $N$-site tight-binding lattice with site-dependent tunneling profile $t_{k}=J\left[1+\lambda\cos(2\pi\beta k+\phi)\right]$. Here $J>0$ denotes the energy scale associated with the tunneling rate and dimensionless $\lambda$ characterizes the strength of the tunneling modulation. When $\beta=1/2$, this model is known as the dimer model or the Su-Schrieffer-Heeger (SSH) model; it describes the transport of charge carriers in acetylene~\cite{ssh1,ssh2}. For a rational value of $\beta$, the one-dimensional AAH model is related to the Hofstadter-butterfly problem~\cite{hoff} which describes the behavior of two-dimensional electron gas in a magnetic field in the presence of a periodic potential. When $\beta$ is irrational, the AAH model describes one-dimensional quasicrystals~\cite{aah1,aah2,qc1,qc2}. For an infinite lattice, when $\beta$ is rational, 
the tunneling amplitude is periodic and the corresponding AAH model has robust topological edge states~\cite{aah3} and is related to topological insulators~\cite{huhughes,esaki,hsopt,bzhu}.  {\it We emphasize that a lattice with tunneling profile $t_k$ is not, in general, reflection-symmetric}~\cite{caveat}. 

In this paper, we investigate the fate of $N$-site AAH models - ones that are experimentally realizable in coupled waveguides or resonators - in the presence of one active, gain potential $+i\gamma$ at site $m_0$ and a balanced loss potential $-i\gamma$ at its reflection-symmetric counterpart site $\bar{m}_0=N+1-m_0$~\cite{yuce}. Our four primary results are as follows. i) When $\beta=1/p$, the model has a positive threshold $\gamma_{PT}(m_0)$ if and only if the lattice size $N$ and the gain location $m_0$ both satisfy $N+1=0\mod p$ and $m_0=0\mod p$. ii) When $\beta=q/p$ is rational, where $p,q$ are co-prime and $q>1$, the same pattern holds irrespective of the value of $q$; when $\beta$ is irrational, the threshold is zero. iii)  When $\beta=q/p$, interspersed among its $p$ bands, the model has $p-1$ localized edge modes that continue to have real energies past the $\mathcal{PT}$-transition. iv) Our predictions are unaffected when the tight-binding lattice approximation is relaxed, and thus are valid in realistic coupled optical waveguides. This work provides a pathway to investigate $\mathcal{PT}$-symmetry breaking in lattice models with topological states. 

The paper is organized as follows. In Sec.~\ref{sec:tbm} we introduce the notation and summarize the properties of a finite, Hermitian AAH model. We present numerical results for the $\mathcal{PT}$-symmetry breaking threshold $\gamma_{PT}(m_0,\phi)$ for a wide range of lattice parameters, and summarize the findings. In Sec.~\ref{sec:hidden}, we present a perturbative analysis of the $\mathcal{PT}$-symmetry breaking threshold and show that, due to a hidden symmetry of the eigenfunctions of the AAH model, the threshold $\gamma_{PT}$ is positive even when though the underlying system is reflection asymmetric. In Sec.~\ref{sec:bpm} we consider the smallest such lattice, a dimer lattice with $N=5$ sites. After an  analytical solution, we present the dynamics obtained via beam-propagation method (BPM), which show that our predictions will remain valid in realistic samples. We conclude the paper with a brief discussion in Sec.~\ref{sec:disc}. 

%---------------------------------------------------------------------------------%

\section{Lattice model and the $\mathcal{PT}$-phase diagram}
\label{sec:tbm}
The Hermitian tunneling Hamiltonian for an $N$-site lattice with nearest neighbor tunneling and open boundary conditions is given by 
\begin{eqnarray}
\label{eq:h0}
H_0(\lambda,\beta,\phi)& = &-\sum_{k=1}^{N-1} t_{k} (|k\rangle\langle k+1|+ |k+1\rangle\langle k|),\\
t_k & = & J\left[1+\lambda\cos(2\pi\beta+\phi)\right],
\end{eqnarray}
where $|k\rangle$ denotes a single-particle state localized at site $k$. The parity (reflection) operator $\mathcal{P}$ on the lattice, in the site-basis, is given by $\mathcal{P}_{ab}=\delta_{a,\bar{b}}$ where $\bar{b}=N+1-b$. The time-reversal operator $\mathcal{T}=*$ where $*$ denotes complex conjugation. The Hamiltonian $H_0(\lambda,\beta,\phi)$ is not, in general, invariant under the $\mathcal{PT}$ operation. The trivial exceptions are a uniform lattice, $\lambda=0$, or a dimer model, $\beta=1/2$, with an even number of lattice sites. 

Since the tunneling function $t_{k}(\lambda,\beta,\phi)$ is periodic in $\beta$ and $\phi$, without loss of generality, we consider $\beta\in[0,1)$ and $\phi\in[0,2\pi)$. It is also straightforward to show that 
\begin{eqnarray}
\label{eq:aahsymmetry}
H_{0}(-\lambda,\beta,\phi) & = & H_{0}(\lambda,\beta,\phi+\pi),\\
H_0(\lambda,1-\beta,\phi) & = & H_0(\lambda,\beta,2\pi-\phi).
\end{eqnarray} 
Therefore, it is sufficient to restrict ourselves to $\lambda\geq 0$ and $\beta\leq 1/2$. The general band structure of the AAH Hamiltonian $H_0(\lambda,\beta,\phi)$ is highly intricate, where the number of bands is determined by $\beta$, and the locations of band degeneracies are determined by $\lambda$ and $\phi$~\cite{hoff,aah1,aah2,qc1,qc2,aah3}. Note that when $\lambda\geq 1$, the tunneling amplitude $t_{k}$ changes sign from positive to negative along the lattice. In addition, for $\lambda\geq 1$ and a rational $\beta=q/p$, the tunneling amplitude vanishes at $k=0\mod p$ and $\phi=\arccos(-1/\lambda)$. For such parameters, the $N$-site chain splits into pieces of size $p$, and the corresponding  Hamiltonian $H_{0}$ becomes block-diagonal. In order to avoid such cases, whose behavior can be trivially understood, we confine ourselves to modulation strengths $0\leq\lambda<1$. In the presence of reflection-symmetric gain-loss potentials $\pm i\gamma$ at sites $m_0$ and $\bar{m}_0$, the lattice Hamiltonian becomes $H=H_{0}+V$ with  
\begin{equation}
\label{eq:v}
V=i\gamma\left( |m_0\rangle\langle m_0| - |\bar{m}_0\rangle\langle\bar{m}_0|\right)=\mathcal{PT}V\mathcal{PT}\neq V^{\dagger}.
\end{equation}

% Schematic of tunneling and band structure.
\begin{figure}[tb]
\centering
\includegraphics[width=\columnwidth]{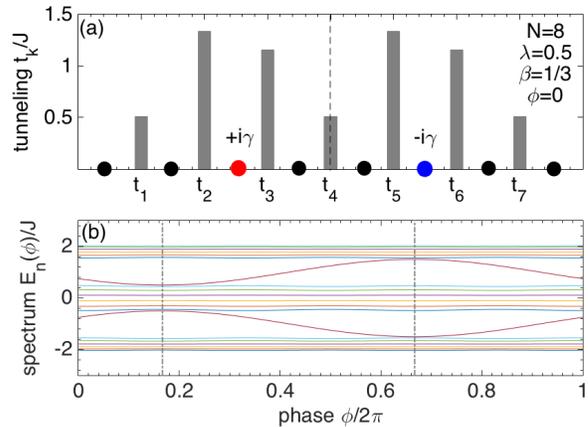}
\caption{(Color online): (a) Schematic of an AAH lattice with $N=8$ sites, denoted by solid circles. The tunneling amplitude $t_{k}/J$ is periodic with period $p=1/\beta=3$. Also shown are balanced gain-loss potentials $\pm i\gamma$, denoted by red and blue solid circles respectively, at reflection-symmetric sites; the vertical dashed line is the lattice center. (b) Spectrum $E_{n}(\phi)$ of an $N=20$ lattice with 
the same periodicity shows $1/\beta=3$ bands, each with $[N\beta]=6$ extended states; [x] denotes the largest integer smaller than x. The two remaining mid-gap states are localized for all $\phi$ except $\phi=\{\pi/3,4\pi/3\}$ shown by dashed vertical lines.}
\label{fig:schematic}
%\vspace{-5mm}
\end{figure}
Figure~\ref{fig:schematic} encapsulates the typical properties of Hamiltonian $H_{0}$. Panel (a) shows the reflection-asymmetrical tunneling profile $t_{k}/J$ for an $N=8$ site lattice with tunneling modulation strength $\lambda=0.5$, inverse tunneling period $\beta=1/3$, and phase $\phi=0$. The neutral sites on the lattice are indicated by solid black circles, the solid red circle at $m_{0}=3$ denotes the gain site, and the loss site at its reflection-symmetric location $\bar{m}_{0}=6$ is denoted by the solid blue circle. Panel (b) shows the energy spectrum $E_{n}(\phi)/J$ for an $N=20$ site AAH model with $\beta=1/3$. In addition to the three bands of extended states that are expected at $\beta=1/3$, there are two edge-localized states with energies that lie in the two band gaps. The midgap states are localized for all values of phases except $\phi=\{\pi/3,4\pi/3\}$, denoted by dotted vertical lines. 

These are generic features of the spectrum for $\beta=q/p$, which corresponds to the tunneling profile  period of $p$, and lattice size $N=Mp-1$. Each of the $p$ bands has $[N\beta]=(M-1)$ extended states, and the remaining $(p-1)$ midgap states are localized for almost all $\phi$. When $N+1=0\mod p$, it is straightforward to show that 
\begin{equation}
\label{eq:pthpt}
\mathcal{PT}H_{0}(\lambda,\beta,\phi)\mathcal{PT}=H_{0}(\lambda,\beta,2\pi\beta-\phi).
\end{equation}
Thus, $H_{0}$ becomes $\mathcal{PT}$ symmetric if and only if $\phi=\{\pi\beta,\pi\beta+\pi\}$. These are precisely the $\phi$-values at which the midgap states become extended. In the following sections, we will see that the topological midgap states do not participate in the $\mathcal{PT}$-symmetry  breaking and retain their localized character past the symmetry breaking transition. 

% Global PT phase diagram. 
\begin{figure}[tpbh] 
\centering
\includegraphics[width=\columnwidth]{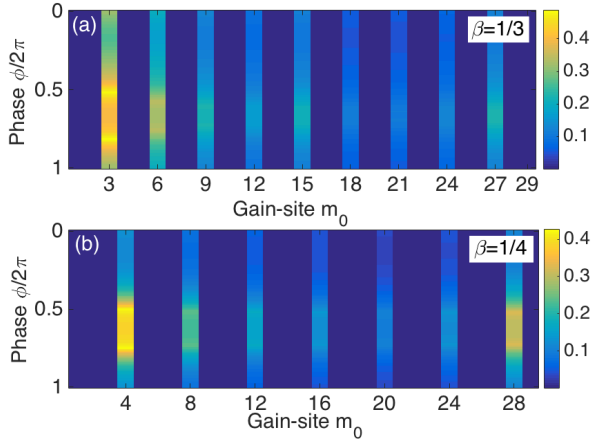}
\caption{(Color online): Threshold $\gamma_{PT}/J$ as a function of gain-potential location $1\leq m_{0}\leq N/2$ and phase $\phi$ for an $N=59$ lattice. (a) When $\beta=1/3$, $\gamma_{PT}=0$ for all gain locations except $m_{0}=\{3,6,\cdots\}$. (b) When $\beta=1/4$, a positive $\gamma_{PT}$ is obtained only when $m_{0}=\{4,8,\cdots\}$. These results show that, contrary to naive expectations, a  reflection-asymmetric AAH Hamiltonian has a positive threshold $\gamma_{PT}(m_{0},\phi)/J\sim 1>0$.}
\label{fig:ptglobal}
%\vspace{-5mm}
\end{figure}

We now present the $\mathcal{PT}$ phase diagram for this model. Naively, the reflection-asymmetric nature of the tunneling Hamiltonian $H_{0}$ will imply, via perturbation theory,  that an infinitesimal gain-loss potential, Eq.(\ref{eq:v}), will lead to a complex spectrum. This expectation is indeed confirmed by numerical results for all lattice configurations except when $\beta=q/p$ is rational, and the lattice size $N$ and gain-site location $m_0$ satisfy $N+1=0\mod p$ and $m_0=0\mod p$. 

Figure~\ref{fig:ptglobal} shows the $\mathcal{PT}$ threshold strength $\gamma_{PT}(m_{0},\phi)$ for a lattice with $N=59$ sites and tunneling modulation strength $\lambda=0.5$. Panel (a) shows the results for a tunneling profile with spatial period $p=3$. {\it The threshold strength is zero except when the gain location is an integer multiple of the tunneling period}, $m_{0}=0\mod 3$. Panel (b) shows that a similar behavior is obtained for $\beta=1/4$. Note that this nonzero threshold results only for periods $p$ such that $N+1=0\mod p$. Thus, for example, when $\beta=q/p=q/7$, the $\mathcal{PT}$-breaking threshold for an $N=59$ site lattice  is identically zero for any $m_0$ and any $q\geq 1$. In general, the nonzero threshold $\gamma_{PT}/J$ first decreases as the gain-potential site $m_{0}$ moves in from the end of the lattice and increases again as it approaches the lattice center, $m_{0}\rightarrow N/2$~\cite{mark}. These results are qualitatively similar for large $N$ and the maximum threshold strength remains the same in the thermodynamic limit, $N\gg 1$. 

\begin{figure*}[] 
\centering
\includegraphics[width=\columnwidth]{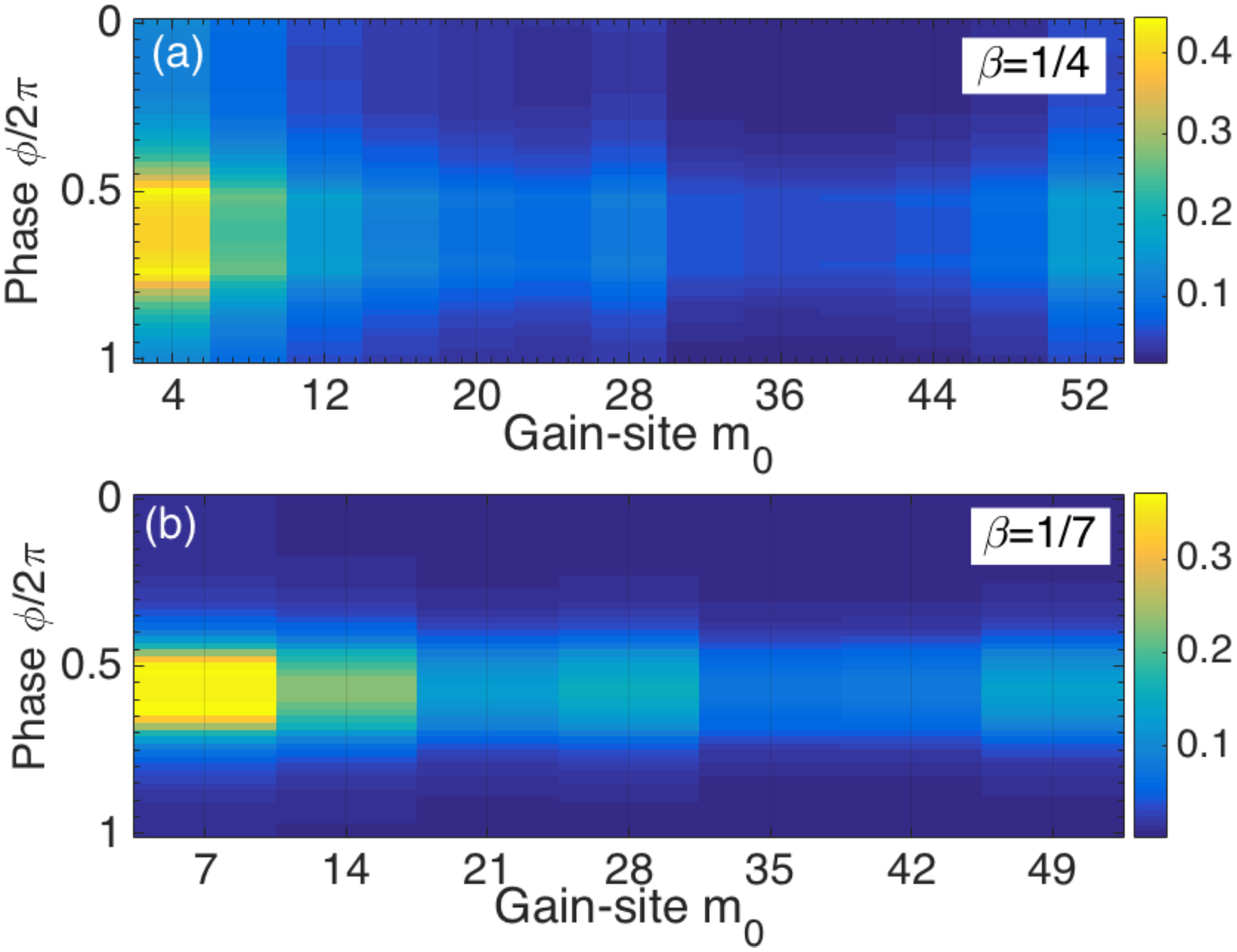}
\includegraphics[width=\columnwidth]{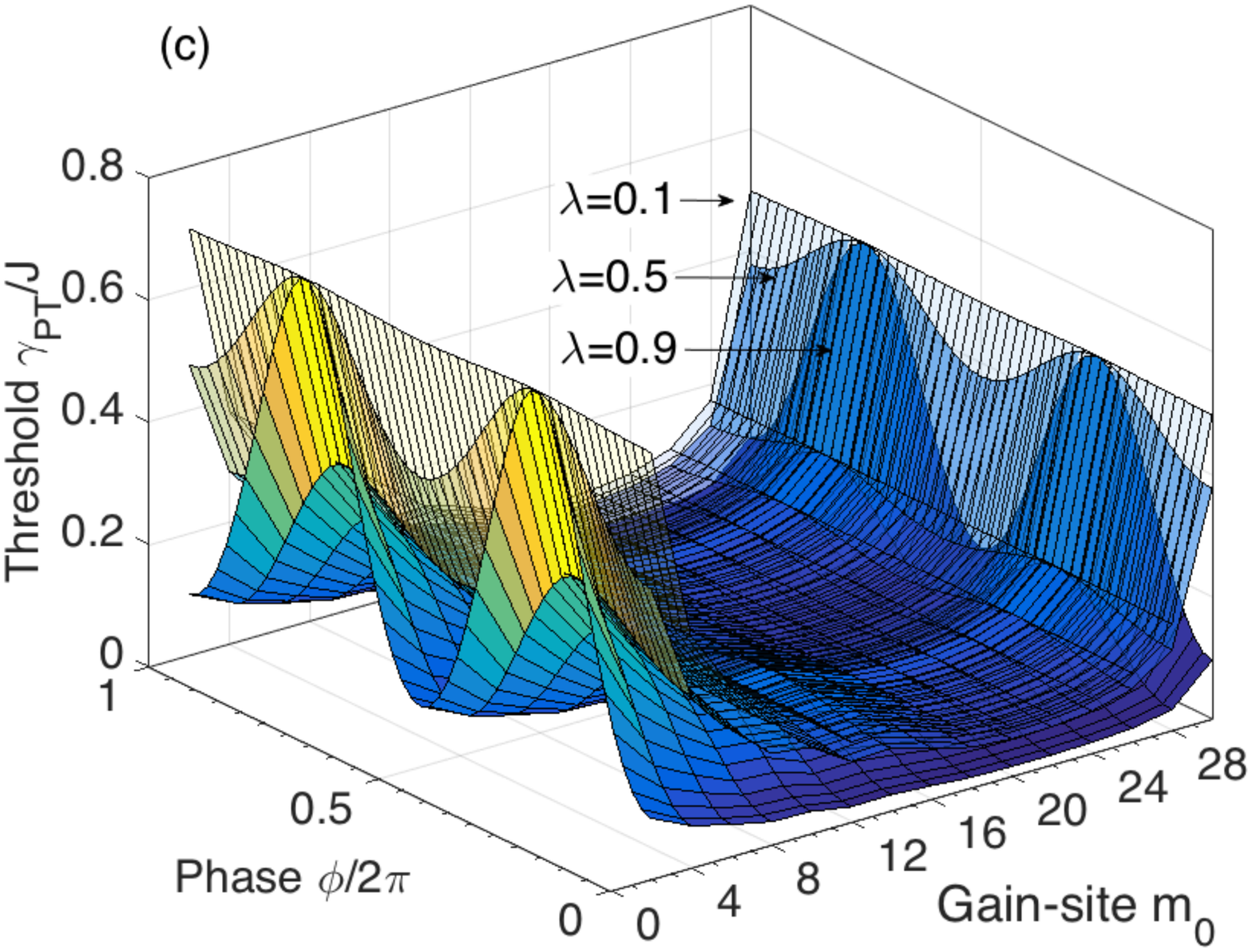}
\caption{(Color online): $\mathcal{PT}$-threshold $\gamma_{PT}(m_{0},\phi)/J$ dependence on the tunneling period $p=1/\beta$ and modulation strength $\lambda$. (a) $\gamma_{PT}(m_{0},\phi)$ for an $N=111$ lattice with $\lambda=0.5$ shows a maximum when gain-loss sites are farthest from ($m_{0}=4$) or closest to ($m_{0}=52$) each other. (b) The same qualitative behavior is observed for the same lattice with a longer tunneling period $1/\beta=7$. (c) Results for an $N=61$ dimer lattice, $\beta=1/2$, show that as the modulation strength $\lambda$ increases, the threshold $\gamma_{PT}$ is monotonically suppressed from its value in the $\lambda\rightarrow 0$ limit~\cite{mark}.}
\label{fig:pt3d}
\end{figure*}

Figure~\ref{fig:pt3d} shows the typical dependence of positive $\gamma_{PT}$ on the tunneling period $p$  and tunneling modulation strength $\lambda$; in each case, only gain-potential locations $m_{0}\leq N/2$ that give rise to a positive $\mathcal{PT}$ threshold are considered. Panels (a) and (b) show the $\mathcal{PT}$ threshold in the $(m_{0},\phi)$ plane for the same modulation strength $\lambda=0.5$ and lattice size $N=111$. Consistent with the results in Fig.~\ref{fig:ptglobal}, the $\mathcal{PT}$ threshold varies non-monotonically with phase $\phi$, and is generally maximum when the gain and loss locations are farthest apart or nearest to each other.  As the tunneling period is increased from $1/\beta=4$, panel (a), to $1/\beta=7$, panel (b), we see that the region with appreciable threshold value shrinks in size but the maximum value of $\gamma_{PT}$ does not alter substantially. 

Panel (c) in Fig.~\ref{fig:pt3d} shows the dependence of the $\mathcal{PT}$ threshold $\gamma_{PT}(m_{0},\phi)$ on tunneling modulation $\lambda$ for a dimer lattice with $N=61$. When $\beta=1/2$, the tunneling amplitude on adjacent bonds alternates between two values $J(1\mp\lambda\cos\phi)$~\cite{ssh1,ssh2}, and the tunneling profile is not reflection-symmetric for an odd $N$. At $\lambda=0.1$, due to the small tunneling modulation, the threshold $\gamma_{PT}$ is essentially independent of the phase $\phi$, and its dependence on $m_{0}$ is similar to that for a uniform tunneling lattice; in particular, we see that $\gamma_{PT}/J\rightarrow 0.5$ when the gain and loss sites are closest to each other~\cite{mark}. As $\lambda$ increases, the $\mathcal{PT}$ threshold, which is proportional to the  effective tunneling amplitude, is strongly suppressed when $\cos\phi=\pm 1$, but remains unchanged from its $\lambda\ll 1$ limit when $|\phi|\approx\pi/2\mod 2\pi$. As an aside, we note for larger tunneling periods $p>2$, the $\phi$-dependence of the threshold $\gamma_{PT}$ is not as easily characterized. Results in Figs.~\ref{fig:ptglobal} and~\ref{fig:pt3d} might suggest that the threshold reaches a maximum at $\phi=\{\pi,\pi+\pi\beta\}$; however, that is not true for all modulation strengths. 

Figs.~\ref{fig:ptglobal} and~\ref{fig:pt3d} capture all global features of the $\mathcal{PT}$ phase diagram. The detailed structure of the $\mathcal{PT}$ threshold manifold $\gamma_{PT}(m_{0},\phi)$ depends on the other two parameters $(\lambda,\beta)$. We emphasize that the {\it $\mathcal{PT}$-symmetry breaking threshold is maximum when the distance between the gain-site and the loss-site is maximum.} Starting from gain-loss sites nearest to each other, $m_0\sim N/2$, we expect that when the distance between them $d=(N+1-2m_0)$ is increased, the $\mathcal{PT}$-symmetry breaking threshold $\gamma_{PT}(d)$ will decrease. Our results, however, predict otherwise. This surprising finding is due to open boundary conditions that ensure complete reflection at the two ends of the lattice. 

What is the origin of the positive threshold $\gamma_{PT}$ when the underlying Hermitian Hamiltonian is not reflection symmetric? In the following section, we present a hidden symmetry of the eigenfunctions of $H$ that is instrumental to a non-vanishing threshold. 

% PT diagram vs lambda and beta. 

%--------------------------------------------------------------------------------% 

% Hidden symmetry and localized state.  
\section{Origin of the positive threshold: hidden symmetry of the AAH model}
\label{sec:hidden}

Let us recall how a positive $\mathcal{PT}$-breaking threshold arises in the case of traditional $\mathcal{PT}$-symmetric Hamiltonians. If the tunneling Hamiltonian is $\mathcal{PT}$-symmetric, so are its eigenfunctions $f_{\alpha}(k)$ with energies $\epsilon_\alpha$. In the presence of a $\mathcal{PT}$ symmetric potential, Eq.(\ref{eq:v}), the first-order perturbative correction to the eigenenergies $\epsilon_\alpha$ is given by 
\begin{equation}
\label{eq:pert}
\Delta_{\alpha}^{1}(\gamma,m_0)= i\gamma(|f_{\alpha}(m_{0})|^{2}-|f_{\alpha}(\bar{m}_{0})|^{2}). 
\end{equation}
Since the eigenfunctions $f_\alpha(k)$ have equal weights on reflection-symmetric sites $(m_0,\bar{m}_0)$,  this correction as well as all higher odd-order corrections vanish for for all gain-locations,  $\Delta_{\alpha}^{2n+1}(\gamma,m_0)=0$ for every $m_0$~\cite{moiseyev}. This property ensures a real spectrum $\epsilon_{\alpha}(\gamma)$ for potential strength $\gamma\leq\gamma_{PT}$. But what if $H_{0}$ is not reflection-symmetric? Its arbitrary eigenstate $f_{\alpha}(k)$ with energy $\epsilon_{\alpha}$ satisfies the following difference equations at reflection-symmetric sites $(k,\bar{k})$.  
\begin{eqnarray}
\label{eq:ef1}
t_{k-1}f_{\alpha}(k-1)+t_{k}f_{\alpha}(k+1) & = & -\epsilon_{\alpha}f_{\alpha}(k),\\
\label{eq:ef2}
t_{\bar{k}-1}f_{\alpha}(\bar{k}-1)+t_{\bar{k}}f_{\alpha}(\bar{k}+1) & = & -\epsilon_{\alpha}f_{\alpha}(\bar{k}),
\end{eqnarray}
where open boundary conditions are implemented by using $t_{0}=0=t_{N}$. It follows that if $t_{k}\neq t_{\bar{k}-1}$, the eigenfunctions, in general, will not have equal weights on the reflection-symmetric sites, $|f_{\alpha}(k)|\neq |f_{\alpha}(\bar{k})|$. 

% Hidden symmetry characterization.
\begin{figure}[h!] 
\centering
\includegraphics[width=\columnwidth]{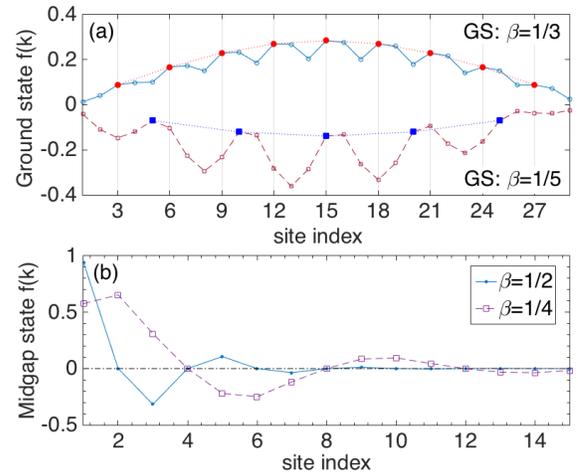}
\caption{(Color online): Hidden symmetry of eigenfunctions of $H_{0}(\lambda,\beta,\phi)$. (a) $N=29$ lattice  with $\lambda=0.5$ shows reflection-asymmetric ground state ($\beta=1/3,\phi=0$: solid line; $\beta=1/5,\phi=\pi$: dashed line). However, the amplitudes on sites $k=0\mod(1/\beta)$ show reflection symmetry about the lattice center (red solid circles, blue solid squares). (b) Lowest energy midgap states for an $N=15$ lattice with $\lambda=0.5$, $\phi=0$, and $\beta=\{1/2,1/4\}$ show that their wave functions vanish at sites $k\propto1/\beta$. These states are unaffected by the balanced gain-loss potential.}
\label{fig:pth}
\vspace{-3mm}
\end{figure}
Figure~\ref{fig:pth} shows typical eigenfunctions of the Hamiltonian $H_{0}(\lambda,\beta,\phi)$ when the lattice size satisfies $N+1=0\mod p$. Panel (a) in Fig.~\ref{fig:pth} shows the ground-state (GS) wave functions $f_G(k)$ for an $N=29$ site lattice with tunneling period $p=3$ (solid line with open circles) and $p=5$ (dashed line with open squares). The results are for tunneling modulation strength $\lambda=0.5$ and phases $\phi=0,\pi$ respectively. As is expected, both ground state profiles are reflection-asymmetric  about the center site $n_{c}=15$. {\it However, these wave functions have the following hidden symmetry}. Solid red circles show the $p=3$ GS amplitudes at sites $k=0\mod p=\{3,6,\ldots\}$,  whereas the solid blue squares show the $p=5$ GS amplitudes at sites $k=0\mod p=\{5,10,\ldots\}$. In both cases the wave function weights satisfy $|f_{\alpha}(m_{0})|=|f_{\alpha}(\bar{m}_{0})|$ if and only if $m_{0}$ is an integer multiple of the tunneling modulation period $p$. This result is true for all eigenstates of $H_{0}(\lambda,\beta,\phi)$ if and only if the lattice size $N$ satisfies $N+1=0\mod p$. It ensures that the reflection-counterpart site index $\bar{m}_{0}=N+1-m_{0}$ is also an integer multiple of the tunneling modulation period. This hidden symmetry is instrumental to a positive $\mathcal{PT}$ threshold that we observe when the gain potential is located at sites $m_{0}=0\mod p$. It also implies, via the perturbation theory arguments~\cite{moiseyev}, that the eigenfunctions of the total Hamiltonian $H=H_{0}+V$ continue to have this symmetry for $\gamma\leq\gamma_{PT}$. 

Next, we consider implications of this hidden symmetry to localized midgap states that are, in some cases, topological in nature. Panel (b) in Fig.~\ref{fig:pth} shows the lowest energy midgap state for an $N=15$ lattice with $\beta=1/2$ (solid line) and $\beta=1/4$ (dashed line); the results are for $\lambda=0.5$ and phase $\phi=0$. These states are localized at one end of the lattice. The surprising feature, shared by all localized midgap states, is the presence of nodes precisely at sites $m_{0}=0\mod p=\{p,2p,\ldots\}$. When the lattice size satisfies $N+1=0\mod p$, due to the tunneling amplitude asymmetry at the two ends of the lattice, it follows that a midgap state must be localized at one end or the other, but not equally at both ends. Therefore, the hidden symmetry discussed in the previous paragraph implies that its wave function must vanish at sites $m_{0}=0\mod p$. This result is true for all $(p-1)$ localized midgap states. This remarkable property of the localized states shows that a balanced gain-loss potential will have no effect on them. In particular, the energies of these states remain real and these localized, topological states~\cite{aah3} remain robust even when the gain-loss strength exceeds the threshold, $\gamma>\gamma_{PT}(\lambda,\beta,\phi;m_{0})$. In recent years, the presence or absence of topological insulator states in $\mathcal{PT}$-symmetric Dirac and SSH models has been extensively studied~\cite{huhughes,esaki,hsopt,bzhu}. Our results show that robust, topological states occur in a wide class of reflection-asymmetric Hamiltonians with a positive $\mathcal{PT}$ breaking threshold. 

%--------------------------------------------------------------------------------% 

\section{Analytical and Beam Propagation Method results: N=5 dimer lattice}
\label{sec:bpm}
The smallest experimental realization of a lattice with reflection asymmetry and a positive $\mathcal{PT}$-breaking threshold, say in a coupled waveguide array, will require $N=5$ waveguides with a dimer tunneling profile $t_1=J(1-\lambda\cos\phi])$ and $t_2=J(1+\lambda\cos\phi)$, and gain-loss potentials $\pm i\gamma$ at reflection-symmetric locations $(m_0,\bar{m}_0)$. This analytically solvable case provides further insights into the results presented in this paper. 

It is easy to check that when $m_{0}=1$, the characteristic equation for the $5\times5$ Hamiltonian $H=H_{0}+V$ has complex coefficients and therefore, the $\mathcal{PT}$ threshold at $m_{0}=1$ is zero. When $m_{0}=2$, the corresponding equation is given by 
\begin{equation}
\label{eq:char}
x\left[x^4-x^2(2t_1^2+2t_2^2-\gamma^2)+( t_1^4+t_2^4+t_1^2t_2^2)\right]=0.
\end{equation}
It follows from Eq.(\ref{eq:char}) that the eigenvalues of $H(\gamma)$ are either purely real or occur in complex conjugate pairs, and the real spectrum has a particle-hole symmetry in the $\mathcal{PT}$-symmetric phase, $\gamma\leq\gamma_{PT}$~\cite{ph}. The threshold gain-loss strength at which the eigenvalues transition from real to complex conjugate pairs is given by 
\begin{equation}
\label{eq:pt5}
\gamma_{PT}(\lambda,\phi) = J\sqrt{2(t_1^2+t_2^2)}\left[1-\sqrt{1-\frac{t_1^2t_2^2}{(t_1^2+t_2^2)^2}} \right]^{1/2}.
\end{equation}
It follows from Eq.(\ref{eq:pt5}) that the threshold $\gamma_{PT}$ is insensitive to the tunneling modulation $\lambda$ when $t_1\approx t_2$, and it is maximally suppressed when $\cos\phi=\pm1$ (see Fig.~\ref{fig:pt3d}(c)). 

It is also straightforward to show that the unnormalized, zero-energy, "edge-state" eigenvector is given by $|f\rangle=( t_{\phi}^{2},0,-t_{\phi},0,1)^{T}$ where $t_{\phi}=t_2/t_1$ is the ratio of tunneling amplitude on the second bond to the tunneling amplitude on the first bond~\cite{jsp}. Thus, the edge state has nodes at sites $k=0\mod 2=\{2,4\}$, is localized at the left end (right end) of the lattice when $t_{\phi}>1$ ($t_{\phi}<1$), and remains unaffected by the $\mathcal{PT}$ potential. 

% BPM results.
\begin{figure*}
\centering
\includegraphics[width=1.9\columnwidth]{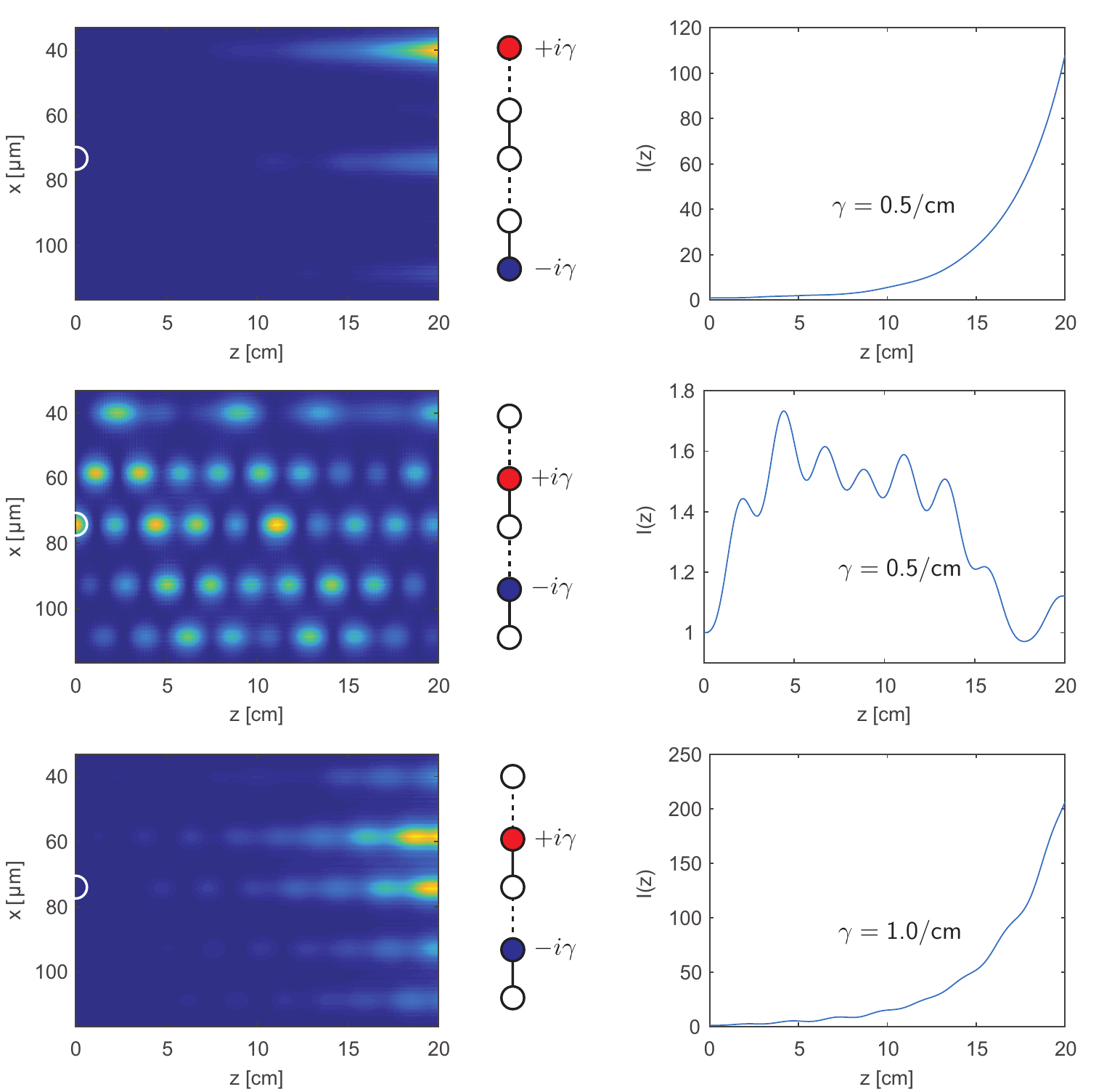}
\caption{(Color Online) BPM results for an array of $N=5$ coupled waveguides. The width of each waveguide is $W_g=5$ $\mu$m and the distances $d\sim10$ $\mu$m between the waveguides are chosen such that the tunneling ratio satisfies $t_1/t_2=3/5$. The left-hand panel in each row shows the space-and-time dependent intensity $I(x,z)$; the initial state is a normalized Gaussian centered on the 3rd waveguide (shown by a white half-circle). The right-hand panel in each row shows the net intensity $I(z)$ as a function of time $t$, or equivalently, the distance $=ct/n_0$ along the waveguide. The center panel shows the schematic of gain and loss locations. (a) The array is in the $\mathcal{PT}$ broken phase when the gain-location is first waveguide. (b) With the same gain-loss strength, the array is in the $\mathcal{PT}$-symmetric phase when $m_0=2$.  (c) When the gain is doubled, the system transitions into the $\mathcal{PT}$ broken phase.}
%\vspace{-3mm}
\label{fig:bpm}
\end{figure*}
Next, we test the validity of our predictions via beam-propagation method (BPM)~\cite{bpm1,bpm2}. This method alleviates the constraint of tight-binding approximation, by taking into account the nonzero spatial dimension of a "single site". With a realistic waveguide-array in mind~\cite{julia}, we obtain the time-evolution of an initially normalized wave packet localized in the center waveguide, $\psi(x,0)=\exp\left[-(x-x_3)^2/4\sigma^2\right]/(2\pi\sigma^2)^{1/4}$. Here $x$ is the continuous coordinate transverse to the waveguide array, $x_3$ is the center of the third waveguide, 
and the initial wave packet size $\sigma$ is set to half-the-width of the waveguide, $\sigma=W_g/2$. We remind the Reader that in this realization, the wave function $\psi(x,t)$ represents the slow-varying envelope of the electric field $E(x,z,t)$ that also has a rapidly varying part proportional to $\exp(ik_0z)$. The time-evolution of $\psi(x,t)$ is given by the Maxwell wave-equation in the paraxial approximation~\cite{bpm1,bpm2}, 
\begin{equation}
\label{eq:bpm}
i\frac{\partial\psi}{\partial t}=-\frac{c}{2k_0n_0^2}\frac{\partial^2\psi}{\partial x^2}+ck_0\left[1-\frac{n(x)^2}{n_0^2}\right]\psi.
\end{equation}
Here $c$ is the speed of light in vacuum, $n_0$ is the cladding index of refraction, $n(x)=n_0+\Delta n(x)$ is the position-dependent index of refraction, and the index contrast $\Delta n\sim 10^{-4}\neq 0$ only within each waveguide. For $\Delta n\ll n_0$, the effective potential is proportional to the index contrast, $V(x)\propto \Delta n$. We implement the gain and loss by adding imaginary parts $\pm i\gamma$ to the real index contrast $\Delta n$.  When the potential $V(x)$ is not real, the time evolution of the wave packet is not unitary, and therefore the total intensity $I(z)=\int dx |\psi(x,t=zn_0/c)|^2$ is not a constant; note that we have switched to the distance along the waveguide $z=c t/n_0$ as a stand-in for the time, for it allows an easier comparison with typical experimental setups.  

Figure~\ref{fig:bpm} shows the results of such an analysis. Each row shows the space- and time-dependent intensity $I(x,z)=|\psi(x,z)|^2$ (left-hand panel); a schematic of the corresponding 5-site lattice (center panel); and the time-dependence of the total intensity $I(z)$ (right-hand panel). The first row shows that when the gain and loss potentials $\pm i\gamma$ are located on the first and the last waveguides respectively, the net intensity $I(z)$ increases monotonically with time, indicating a $\mathcal{PT}$-symmetry broken phase. The second row shows the results when the gain-loss are at sites $m_0=2$ and $\bar{m}_0=4$ respectively. It is clear from the $I(x,z)$ plot that the wave packet undergoes oscillations across the lattice along with some amplification. This periodic behavior is also manifest in the total intensity $I(z)$ and shows that the system is in the $\mathcal{PT}$-symmetric phase for the same gain-loss strength. The bottom row shows that when the gain-loss strength is doubled, the system enters $\mathcal{PT}$-broken phase, as evidenced by monotonically increasing net intensity $I(z)$. 

These results demonstrate that the non-trivial dependence of the $\mathcal{PT}$-breaking threshold on the gain location $m_0$ for reflection-asymmetric models is robust. 
%--------------------------------------------------------------------------------% 

\section{Discussion}
\label{sec:disc}

In this paper, we have discovered that a broad class of Aubry-Andre Harper models~\cite{aah1,aah3} with reflection-asymmetric tunneling profile can have a positive $\mathcal{PT}$-symmetry breaking threshold. This occurs when $\beta=q/p$ is rational, and the lattice size $N$ and the gain-potential location $m_0\leq N/2$ both satisfy $N+1=0\mod p$ and $m_0=0\mod p$. These constraints ensure that the loss-potential location $\bar{m}$ also satisfies $\bar{m}=0\mod p$. Through the tight-binding analysis of the lattice model and a BPM analysis of its continuum counterpart, we have shown that our predictions remain valid for realistic waveguide arrays. The AAH lattice models investigated here are known to support topological states~\cite{aah3}. They thus provide an avenue to experiments in which the interplay between $\mathcal{PT}$-symmetry breaking and topological properties can be studied. 

We note that this paper is based on an effective, single-particle Hamiltonian that permits amplification and decay. {\it Prima facie}, these results predict the existence of topological insulators with positive $\mathcal{PT}$ breaking threshold~\cite{huhughes,esaki}, since our model makes no reference to the quantum statistics of the particle. In reality, however, amplification of a single degree of freedom is incompatible with the Pauli principle. Thus, our results can apply to fermions only if the gain and loss are associated with the bulk Fermi sea, and not with a single quantum degree of freedom. In the bosonic case, amplification of a single degree of freedom is permitted and our results are directly applicable. We have also ignored two-body interactions; they become important only in the $\mathcal{PT}$-broken phase as the on-site intensity (light) or density (massive bosons) is amplified. 

%--------------------------------------------------------------------------------%
\section*{acknowledgment}
This work is supported by NSF DMR-1054020. 
%---------------------------------------------------------------------------------%

%---------------------------------------------------------------------------------%

\begin{thebibliography}{99}
\bibitem{caveat} The tunneling profile is reflection-symmetric, $t_k=t_{N-k}$, if and only if $\phi=\pi\beta N\mod 2\pi$. We ignore such accidental cases, since they form a set of measure zero.
\end{thebibliography}
\end{document}